\documentclass[fleqn,12pt,twoside]{article}
\usepackage[headings]{espcrc1}
\usepackage{bm}

\readRCS
$Id: espcrc1.tex,v 1.2 2004/02/24 11:22:11 spepping Exp $
\usepackage{graphicx}
\usepackage[figuresright]{rotating}

\newcommand{\AmS}{{\protect\the\textfont2
  A\kern-.1667em\lower.5ex\hbox{M}\kern-.125emS}}

\title{Isotropization by QCD Plasma Instabilities}

\author{Yasushi Nara\address[FRK]{Institut f\"ur Theoretische Physik,
          Johann Wolfgang Goethe Universit\"at,\\
	  Max von Laue Str.\ 1, D-60438 Frankfurt am Main, Germany}%
	  }
       
\begin{document}

\maketitle

\begin{abstract}
Numerical solutions of the Wong-Yang-Mills equations
with anisotropic particle momentum distributions
are presented. Their isotropization by collective
effects due to the classical Yang-Mills field is shown.
\end{abstract}

\section{INTRODUCTION}

It is of great importance to understand the non-equilibrium dynamics
in the early stage of high energy heavy-ion collisions and if, how,
and when the produced partons thermalize.  Boltzmann transport
approaches with inelastic processes have been employed to follow the
scattering of the hard gluons among themselves~\cite{PCM}.  However,
the soft classical color fields may give rise to non-perturbative
collective processes and hence could also play an important role.
Specifically, QCD plasma instabilities may develop, e.g.\ due to anisotropic
distributions of released hard partons~\cite{Mrowczynski}.  As a
consequence, the ``bottom-up scenario''~\cite{BMSS} for thermalization
of the Color Glass Condensate would be modified significantly~\cite{ALM}.

Indeed, within the hard loop effective theory unstable soft modes were
found~\cite{HTL_unst_mode} if the hard modes exhibit an anisotropic
distribution in momentum space. The time evolution has been studied
via the full non-linear hard loop effective
action~\cite{Rebhan:2004ur}.
$3+1$D simulations of that theory~\cite{AMY}
show that instabilities grow more slowly once the field strength
becomes large and
non-Abelian self-interactions become nonperturbative.
Possible effective potentials
for anisotropic QCD plasmas beyond the hard loop approximation
are discussed in Ref.~\cite{MM}.
 
In the present work we follow numerically the evolution of the
hard modes, represented by particles, coupled to a classical SU(2)
Yang-Mills field. The back-reaction of the fields on the particles is
taken into account fully. We present some additional results as compared to
ref.~\cite{DN05}, such as the modification of the particle spectra.

\section{Model}

We solve Wong's equations~\cite{Wong}
\begin{equation}
\frac{d\bm{x}_i}{dt}  =  \bm{v}_i,\qquad
\frac{d\bm{p}_i}{dt}  =  gQ_i^a \left( \bm{E}^a + \bm{v}_i \times
  \bm{B}^a \right),\label{pdot}\qquad
\frac{dQ_i}{dt}  =  igv^{\mu}_i [ A_\mu, Q_i],
\end{equation}
for the $i$-th (test) particle, coupled to the Yang-Mills equation
\begin{equation}
 D_\mu F^{\mu\nu} = j^\nu 
 = g \sum_i Q_i v^\nu \delta(\bm{x}-\bm{x}_i).
\end{equation}
These equations
reproduce the ``hard thermal loop'' effective
action~\cite{ClTransp} near equilibrium.
Numerical techniques to solve the classical field equations coupled to
particles have been developed in Ref.~\cite{HuMullerMoore}.
Our update algorithm is closely related to the one explained there.

In the following, we assume that the fields only depend on time and on
one spatial coordinate, $x$, which reduces the Yang-Mills equations to
1+1 dimensions. The particles are allowed to propagate in three
spatial dimensions.  For simplicity, we also restrict ourselves to the
case without expansion here; the more realistic case with longitudinal
expansion (solved without particles in~\cite{KV}) will be addressed in
the future.
The initial anisotropic phase-space distribution of hard gluons is taken to be
$
  f(\bm{p},\bm{x})\propto \exp(-\sqrt{p_y^2+p_z^2}/p_\mathrm{hard})\,
\delta(p_x)~.
$
This represents a quasi-thermal distribution in two dimensions, with
``temperature'' $=p_\mathrm{hard}$. It is assumed to come about by free
streaming of the particles for a time of order
$\sim1/p_\mathrm{hard}$ after the collision of the nuclei. In a
comoving frame (with the central rapidity region) then, the longitudinal
momenta are red-shifted to much below the typical transverse momentum.

The initial field amplitudes are sampled from a Gaussian distribution
with a width tuned to a given initial energy density.
We solve the Yang-Mills equations in $A^0=0$ gauge and also set
$\bm{A}=0$ (i.e.\ all gauge links =1) at time $t=0$; the initial electric
field is taken to be polarized in a random direction transverse to the x-axis.

\section{RESULTS}

\begin{figure}[htb]
\begin{minipage}[t]{80mm}
\includegraphics[width=3in]{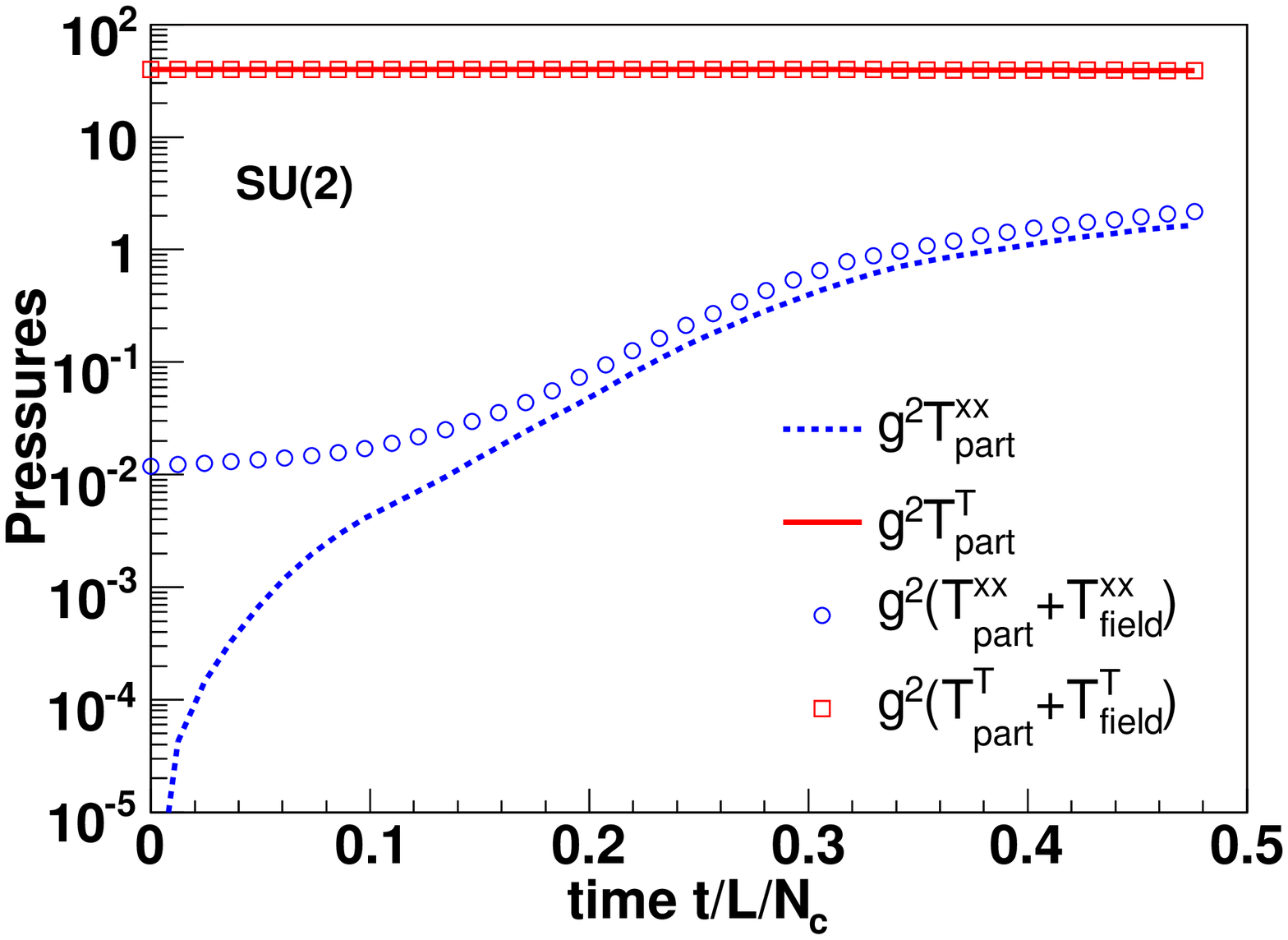}
\caption{Time evolution of the pressures for weak initial fields.}
\label{fig:tmnw}
\end{minipage}
\hspace{\fill}
\begin{minipage}[t]{75mm}
\includegraphics[width=3in]{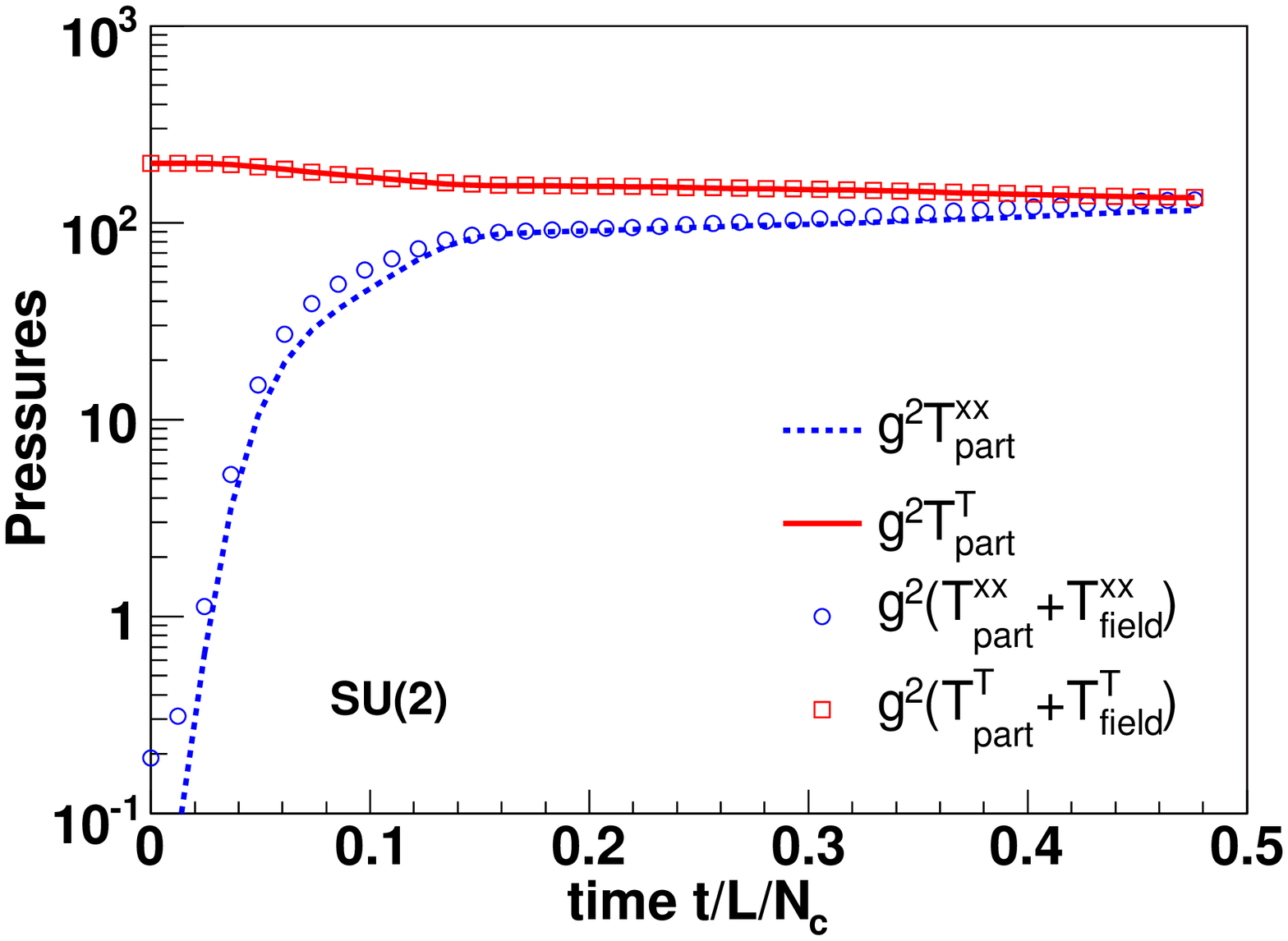}
\caption{Time evolution of the pressures for the strong field case.}
\label{fig:tmns}
\end{minipage}
\end{figure}
Fig.~\ref{fig:tmnw} shows the time evolution of the particle and
field pressures for the initial condition
$p_{\mathrm{hard}}=10$~GeV and initial field energy density
$\approx 10^{-2}$ GeV/fm$^3/g^2$, for which instabilities have been seen
in Ref.~\cite{DN05}.
We observe a rapid exponential growth of the longitudinal pressure,
approaching an isotropic configuration. This is due to deflection of
the particles in the exponentially growing field, which
randomizes their momenta. Note also that the dominant contribution to
the build-up of longitudinal pressure is from deflection of the
particles, not from the fields themselves. In our
1+1D simulations the transverse pressure of the fields is zero
$T^{T}_{\mathrm{field}}=0$.  Furthermore, we find that
$T^{00}_{\mathrm{field}}=T^{xx}_{\mathrm{field}}$, i.e.\ equal energy
density and pressure, which is due to the
dominance of transverse magnetic fields.  Since periodic boundary
conditions are used, no collective flow is generated: $T^{0i}=0$.

The separation between hard and soft modes may not be very large at
RHIC energy~\cite{KV}.  In Fig~\ref{fig:tmns}, we show the time evolution of
the pressures for $p_{\mathrm{hard}}=1$ GeV and initial field energy
density $\approx 10^{-1}$ GeV/fm$^3/g^2$; the time scale is set by the
lattice size $L$, =10~fm for this simulation.  When the separation between
hard and soft modes is small, strong instabilities are not
seen, but the system still approaches isotropy~\cite{DN05}. In fact,
the rapid isotropization of the particle momenta by the strong fields
prevents the occurrence of pronounced instabilities. 

\begin{figure}[htb]
\begin{minipage}[t]{80mm}
\includegraphics[width=3in]{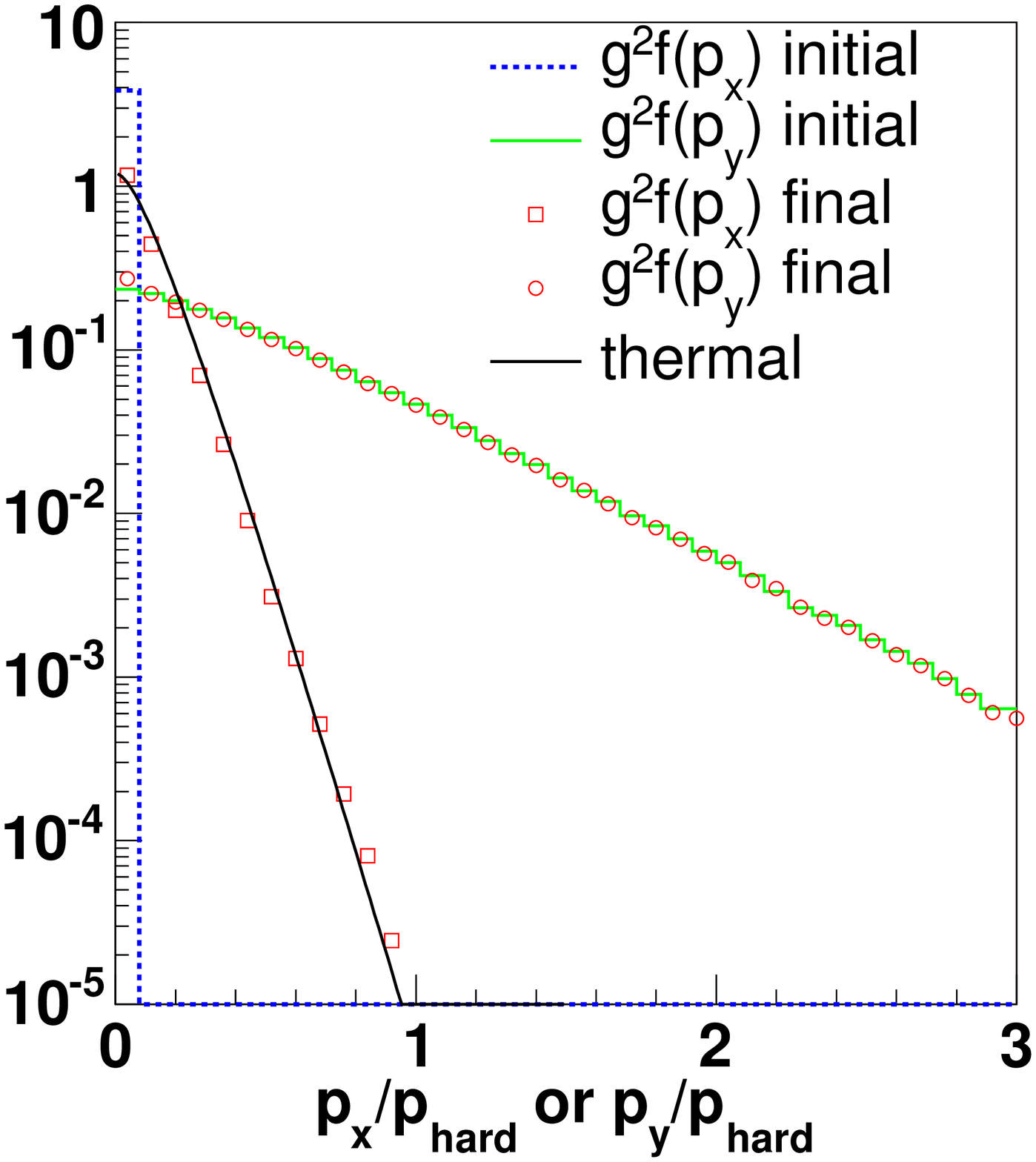}
\caption{Particle distribution functions for the weak field case.}
\label{fig:ptw}
\end{minipage}
\hspace{\fill}
\begin{minipage}[t]{75mm}
\includegraphics[width=3in]{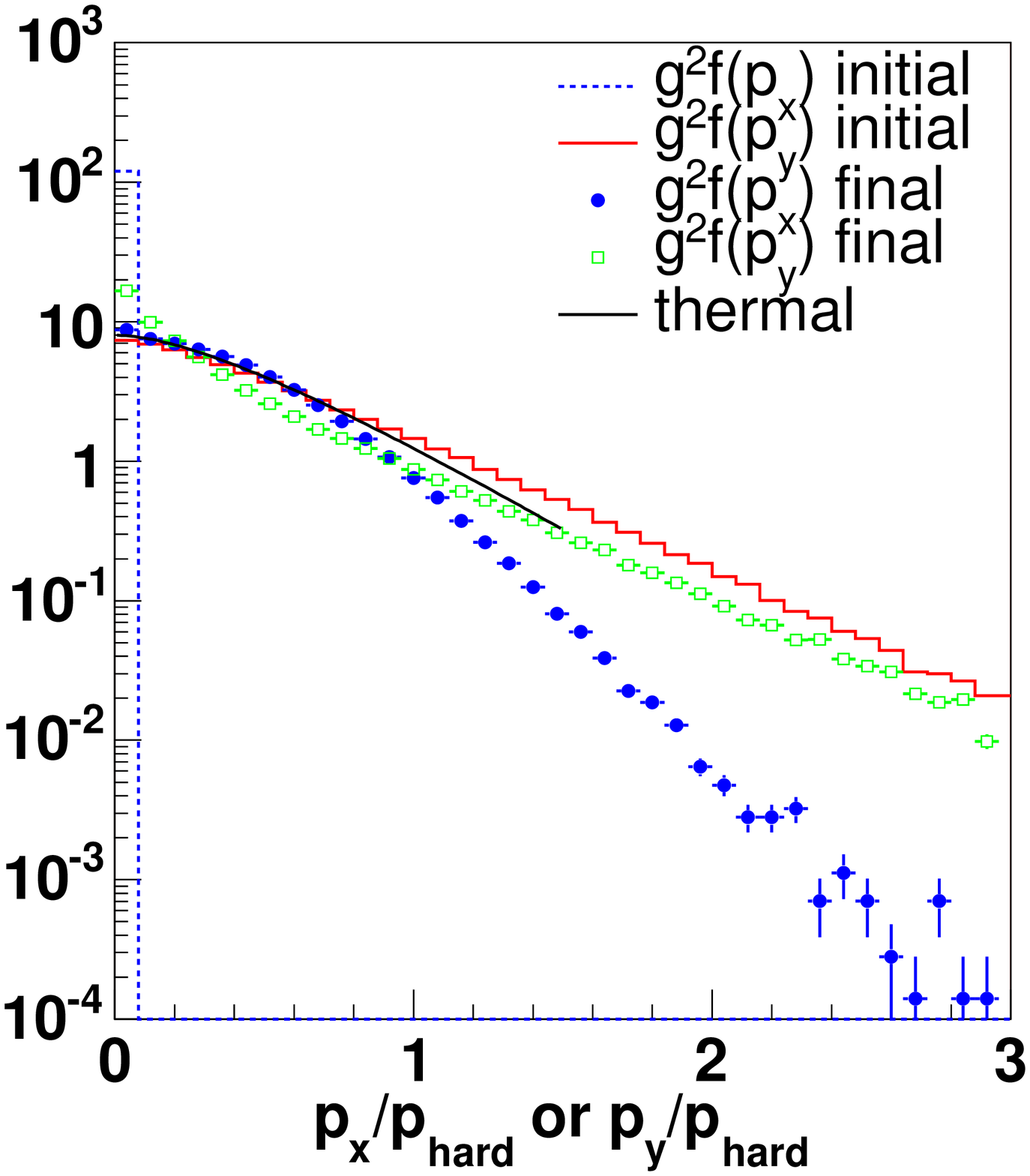}
\caption{Particle distribution functions for the strong field case.}
\label{fig:pts}
\end{minipage}
\end{figure}
It is interesting to look at the particle distribution function
itself, even though we presently restrict ourselves to simulations of
collisionless plasmas.  Field fluctuations generate an
effective collision term~\cite{flucs} mediating soft exchanges.  We
plot initial as well as final ($t/L/N_c=0.5$)
particle distribution functions for weak
initial field in Fig.~\ref{fig:ptw}, and for strong field in
Fig.~\ref{fig:pts}. For the former case, we observe that the initial
$\delta$-function in the longitudinal direction becomes a thermal
distribution extending over several orders of magnitude. However, the
longitudinal direction is still much colder than the transverse
directions, in agreement with the behavior of the pressures above.
This is because of the assumption of transversally homogeneous fields.
The fluctuations of the gauge field induce fluctuations in the motion
of the quasiparticles.  The presence of fluctuations is closely linked
to dissipative processes. 

In the future, full 3D simulations of the Wong-Yang-Mills equations
are required. They will also allow us to study the effects due to a
(azimuthally asymmetric) boundary. We also need to include
longitudinal expansion of the system to check whether the distribution
{\em stays} isotropic. Finally, collisions with momentum transfer
above that mediated by the classical field have to be included for a complete
description of the non-perturbative plasma in weak coupling.

\section*{ACKNOWLEDGMENTS}
The author would like to thank Adrian Dumitru for collaboration
and for carefully reading this manuscript prior to publication.

\end{document}